\documentclass[twocolumn,showpacs,prl,amsmath,amssymb,superscriptaddress]{revtex4}
\usepackage{amsmath,graphicx,bm,amssymb,color}

\newcommand{\Fref}[1]{Fig.~\ref{#1}}
\newcommand{\nin}{\in \hspace{-.37em}/ }
\newcommand{\ninn}{\in \hspace{-.90em}/ }
\newcommand{\figwidth}{2.75in}

\begin{document}

\title{Funnel landscape and mutational robustness as a result of
evolution under thermal noise}
 
\author{Ayaka Sakata}
\affiliation{Graduate school of Arts and Sciences, The University of
Tokyo, Komaba, Meguro-ku, Tokyo 153-8902, Japan}
\author{Koji Hukushima}
\affiliation{Graduate school of Arts and Sciences, The University of
Tokyo, Komaba, Meguro-ku, Tokyo 153-8902, Japan}
\author{Kunihiko Kaneko}
\affiliation{Graduate school of Arts and Sciences, The University of
Tokyo, Komaba, Meguro-ku, Tokyo 153-8902, Japan}
\affiliation
{Complex Systems Biology Project, ERATO, JST, Tokyo, Japan}

\begin{abstract} 
Using a statistical-mechanical model of spins,
evolution of phenotype dynamics is studied.
Configurations of spins and their interaction $\bm{J}$
represent phenotype and genotype, respectively. 
The fitness for selection of $\bm{J}$ is given by
the equilibrium spin configurations determined by a
Hamiltonian with $\bm{J}$ under thermal noise. The genotype $\bm{J}$ evolves
through mutational changes under selection pressure  to raise its
fitness value.
From Monte Carlo simulations 
we have found that the frustration around the target spins disappears 
for $\bm{J}$ evolved under temperature beyond a certain threshold. 
The evolved $\bm{J}$s give the funnel-like dynamics, 
which is robust to noise and also to mutation.
\end{abstract}

\date{\today}
\maketitle


Under fixed conditions, biological systems evolve to increase their fitness, determined by a biological state -phenotype- that is shaped by a dynamical process. This dynamics is generally stochastic as they are subject to thermal noise, and
the rule for the dynamics is controlled by a gene that mutates through generations. Those genes that produce higher fitness values have a higher chance of survival.
For example, the folding dynamics of a protein or t-RNA shapes a structure under thermal noise to produce a biological function, while the rule for the dynamics is coded by a sequence of DNA.

The phenotype that provides fitness is expected to be rather insensitive
to the type of noise encountered during the dynamical process \cite{Waddington,Wagner-book,Alon}, to continue producing fitted phenotypes. 
The first issue is to determine
what type of dynamics is shaped through evolution to achieve robustness to noise.
To have such robustness
it is preferable to utilize a dynamical process that produces and maintains target phenotypes of high fitness smoothly and globally
from a variety of initial configurations.
In fact, the existence of such a global attraction 
in the protein folding was proposed as a
consistency principle\cite{Go} and
a 
\textit{funnel} 
landscape\cite{Onuchic}, while similar global attraction has been
discovered in gene regulatory networks\cite{FangLi-Ouyang-Tang}.  
Despite the ubiquity of such funnel 
landscapes for phenotype dynamics,
little is understood on how these structures are shaped by the evolution
\cite{SY,Ancel-Fontana,Deem}. The second issue we address is 
the conditions under which funnel-like dynamics evolves.

Besides its robustness to noise, the phenotype is also expected to be robust against mutations in the genetic sequence encountered through evolution. 
Despite recent studies suggesting
a relationship between robustness to thermal noise and robustness to 
mutation\cite{Waddington,Wagner,KK-book,KK-PLoS,KK-chaos},
a theoretical understanding of the evolution of the two is still insufficient. 
The question of whether robustness to thermal noise also leads to robustness to mutation constitutes the third issue discussed in this paper. All these questions concerning robustness to noise and mutation or the shaping of the funnel-like landscape can be answered by introducing an abstract statistical-mechanical model of interacting spins, whose Hamiltonian evolves over generations to achieve a higher fitness.

Let us consider a system of $N$ Ising spins interacting globally. 
In this model, configurations of spin variables $S_i$ and an interaction
matrix $J_{ij}$ with $i,j=1,\cdots,N$ represent phenotype and
genotype, respectively. 
For simplicity, $J_{ij}$ is restricted to the two values $\pm 1$ and 
is assumed to be symmetric: $J_{ij}=J_{ji}$. 
The dynamics of the phenotype denoted by $\bm{S}$ is given by a 
flip-flop update of each spin with an energy function, defined
by the Hamiltonian $H(\bm{S}|\bm{J})=-\frac{1}{\sqrt{N}}\sum_{i<j}J_{ij}S_iS_j$,
for a given set of genotypes denoted by $\bm{J}$. 
We adopt the Glauber dynamics as an update rule, where the $N$ 
spins are in contact with a heat bath 
of temperature $T_S$.
After the relaxation process, this dynamics
yields an equilibrium distribution for a given $\bm{J}$, 
$P(\bm{S}|\bm{J},T_S)=e^{-\beta_SH(\bm{S}\mid\bm{J})}\slash Z_S(T_S)$,
where $\beta_S=1\slash T_S$ and
$Z_S(T_S)=\mbox{Tr}_{\bm{S}}\exp[-\beta_SH(\bm{S}|\bm{J})]$. 

The genotype $\bm{J}$ is transmitted to the next generation with some
variation, 
whereas genotypes that produce a phenotype with a higher level of fitness are selected.
We assume that fitness is a function of the configuration of target spins $\bm{t}$,
a given subset of $\bm{S}$, with size $t$.
The time-scale for genotypic change is generally much larger than that
for the phenotypic dynamics,  
so that the variables $\bm{S}$ are well
equilibrated within the unit time-scale of the slower variable $\bm{J}$. 
Such being the case, fitness
should be expressed as a function of the target spins $\bm{S}$ 
averaged with respect to the distribution. Considering the gauge transformation \cite{Nishimori},
we define fitness as 
$$
\Psi(\bm{J}|T_S)=\Big\langle\prod_{i<j\in\bm{t}}\delta(S_i-S_j)\Big\rangle, 
\label{Fit}
$$
without losing generality, where $\langle\cdots\rangle$ denotes the expectation value 
with respect to the equilibrium probability distribution. 
In other words, the 
expected fitness is given by the probability of 
the ``target configuration'' in which
all target spins are aligned parallel under equilibrium conditions.
Note that in our model, only the 
target spins contribute
to the fitness,
and as a result, the spin configuration for a given fitness value has redundancy.

The genotype $\bm{J}$ evolves as a result of mutations, random flip-flop of the matrix element,
and the process of selection according to the fitness function. 
We again adopt Glauber dynamics by using fitness 
instead of the Hamiltonian in the phenotype dynamics, 
where $\bm{J}$ is in contact with a heat bath whose 
temperature $T_J$ is different from $T_S$.
In particular, the dynamics is given by a stochastic Markov process with
the stationary distribution 
$P(\bm{J},T_S,T_J)=e^{\beta_J\Psi(\bm{J},T_S)}\slash Z_J(T_S,T_J)$,
where $\beta_J=1/T_J$ and $Z_J(T_S,T_J)=\mbox{Tr}_{\bm{J}}\exp[\beta_J\Psi(\bm{J},T_S)]$.
According to the dynamics, genotypes
are selected somewhat uniformly at high 
temperatures $T_J$, 
whereas at low $T_J$, genotypes with higher fitness values are
preferred.
The temperature $T_J$ represents 
selection pressure.

Next, we study the dependence of the fitness and energy on $T_S$ and $T_J$,
given by 
$\Psi(T_S,T_J)=\langle\Psi(\bm{J}|T_S)\rangle_J$ 
and $E(T_S,T_J)=\langle\langle H(\bm{S}|\bm{J})\rangle\rangle_J$,
respectively, where $\langle\cdots\rangle_J$ denotes the average 
with respect to the equilibrium probability distribution, $P(\bm{J},T_S,T_J)$.
For the spin dynamics (unless otherwise mentioned), the exchange Monte Carlo (EMC) simulation \cite{EMC} 
is used to accelerate 
the relaxation to equilibrium.
Indeed, we have confirmed the equilibrium distribution for the simulations below.
Two processes are carried out alternately:
the equilibration of $\bm{S}$ with the EMC
and the stochastic selection of $\bm{J}$
according to the fitness value estimated through the first process.

\begin{figure}
\includegraphics[width=3.375in]{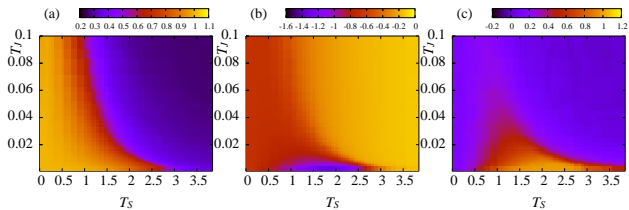}
\caption{(color online)
Contour maps on $T_S$ and $T_J$ of (a) fitness, (b) energy, and (c) $\Phi_2$, where $1-\Phi_2$ is the frustration around the target spins,
for evolved $\bm{J}$ at a given $T_S$ and $T_J$ (see text for details). $N=15$ and $t=3$. 
For each generation of the genotype dynamics, the average 
in equilibrium 
is taken
over 1500 MC steps after discarding the first 1500 
MC steps, which are sufficient for equilibration.
The data are averaged over the last $10^3$ generations.}
\label{Fit_contour}
\end{figure}
 
\Fref{Fit_contour} (a) and (b) show
dependence of the fitness and energy on $T_S$ and $T_J$, respectively, 
for $N=15$ and $t=3$. 
For any $T_S$, the fitness value decreases monotonically with
$T_J$. 
However, $T_S$ influences the slope of the decrease significantly. The
fitness for sufficiently low $T_S$ remains at a high level and 
decreases only slightly with an increase in $T_J$, while for a medium value of $T_S$, the
fitness gradually falls to a lower level as a function of $T_J$ and
eventually, for a sufficiently high value of $T_S$, it never attains a high level. 
This implies that the structure of the fitness landscape 
depends on $T_S$, 
at which the system has evolved. 
The energy function, on the other hand, shows a significant dependence on $T_S$.
Although the energy increases monotonically with $T_S$ for high $T_J$, 
it exhibits non-monotonic behavior at a low $T_J$ and
takes a minimum at 
an intermediate $T_S$. 
The $\bm{J}$ configurations giving rise to the highest fitness value generally have
a 
large redundancy. At around $T_S\simeq 2.0$,
using a fluctuation induced by $T_S$, a specific subset of 
adapted $\bm{J}$'s 
giving rise to lower energy is selected from the
redundant configurations with higher fitness. 

In the medium-temperature range, such $\bm{J}$s that yield both lower energy and 
higher fitness are evolved. Now we study the characteristics of such $\bm{J}$s.
According to statistical physics of spin systems,
triplets of interactions that satisfy $J_{ij}J_{jk}J_{ki}<0$ are known to yield
frustration, which is an obstacle to attaining the unique global energy minimum \cite{Nishimori}.
In our model, however, the target spins play a distinct role.
Hence, it becomes necessary to quantify 
the frustration by distinguishing target and non-target spins. 
In accordance with the ``ferromagnetic'' fitness condition for target spins, we define $\Phi_1$ as the frequency of positive coupling among target spins, 
$
\Phi_1(T_S,T_J)=\frac{2}{t(t-1)}\Big\langle\sum_{i<j\in\bm{t}}J_{ij}\Big\rangle_J\label{J1}
$. 
Under ferromagnetic coupling,
the target configurations are energetically favored, 
i.e., $\Phi_1=1$, for which no frustration exists
among the target spins.
Next, for a measure of the degrees of frustration 
between target and non-target spins,
and among non-target spins themselves, we define $\Phi_2$ and $\Phi_3$ as
$$
\Phi_2(T_S,T_J)=\frac{2}{t(t-1)(N-t)}\Big\langle\sum_{i<j\in\bm{t}}\sum_{k\nin\bm{t}}J_{ik}J_{kj}\Big\rangle_J,
 \label{J2}
$$
and 
$$
\Phi_3(T_S,T_J)=\frac{1}{C^{N-t}_2}\Big\langle\sum_{k<l\nin\bm{t}}\Big(\frac{1}{t}\sum_{i\in\bm{t}}J_{ik}\Big)J_{kl}\Big(\frac{1}{t}\sum_{j\in\bm{t}}J_{lj}\Big)\Big\rangle_J,\label{J3}
$$
where $C^{N-t}_2$ is the total number of possible pairs among the non-target spins.
Here, $\Phi_2$ is the fraction of the interaction pairs between target and non-target spins
that satisfy $J_{ik}
 J_{kj}=1\ (i,j\in\bm{t},\ k\ninn\ \bm{t})$. 
If $\Phi_2=1$, no frustration is introduced by the interaction between target and non-target spins,
so that the energy minimum of the target configuration is conserved by 
such interaction.
If $\Phi_3=1$, 
the target configurations do not introduce frustration in $J_{kl}$s $(k,l\ninn\ \bm{t})$.
If $\Phi_1=\Phi_2=\Phi_3=1$, there is no frustration at all over the interactions,
as suggested by the Mattis model \cite{Mattis}, which can be transformed into
ferromagnetic interactions by gauge transformation \cite{Nishimori}.

\begin{figure}
\includegraphics[width=\figwidth]{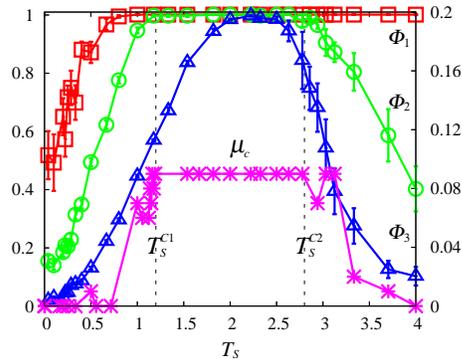}
\caption{(color online) 
Dependence of local frustrations on $T_S$,; $\Phi_1(\square), \Phi_2(\bigcirc), \Phi_3(\triangle)$ 
represents the left axis and $\mu_c(\ast)$ represents the right axis 
by fixing $T_J$ at $0.5\times 10^{-3}$. 
The data are computed by taking averages over 
150 genotypes $\bm{J}$ evolved at a given temperature $T_S$.
$\mu_c$ is a threshold value, 
beyond which the fitness of the mutated genotype begins to decrease (see \Fref{Fit_vs_mu}). The transition points 
$T_S^{c1}$ and $T_S^{c2}$ are estimated 
as a temperature at which $\Phi_2$ deviates from 1.
} 
\label{mu_c}
\end{figure}

For 
$\bm{J}$ that is evolved under given $T_S$ and $T_J$, we have computed 
$\Phi_1,\Phi_2$ and $\Phi_3$. \Fref{mu_c} shows the dependence of $\Phi_1,\Phi_2$ and $\Phi_3$ on $T_S$ 
at a fixed $T_J=0.5\times10^{-3}$.
For $T_S\geq T_S^{c1}$, $\Phi_1$ takes the value $\sim
1$ \cite{remark}, 
so that a target configuration is embedded as an energetically favorable
state, 
while no specific patterns, apart from the target, are embedded in the spin configuration.

For $T_S^{c1}\leq T_S\leq T_S^{c2}$, $\Phi_2$ also takes the value
around 
1, implying that frustration
is not introduced by means of interactions with a non-target spin.
In this temperature range, $\Phi_3$ is not equal to 1, except for $T_S\sim 2.0$ where the Mattis state is shaped.
When $\Phi_2 \sim 1$ and $\Phi_3\neq 1$, frustration is not completely eliminated 
from the non-target spin interactions, 
in contrast to the Mattis state. 
Here, such a $\bm{J}$ configuration
without frustration around the target spins (but with
frustration between non-target spins) is referred to as ``local Mattis
state'' (LMS), 
as characterized
by $\Phi_1=\Phi_2\sim 1$ and $\Phi_3\neq 1$.
The interactions $\bm{J}$ that form such LMSs arise 
as a result of the evolution 
at $T_S^{c1}\leq T_S\leq T_S^{c2}$, where both a fitted target configuration and 
a lower energy level are achieved.
As shown in \Fref{Fit_contour}(c), the $T_S$ range
in which LMS is shaped becomes
narrower with an increase in $T_J$\cite{SHK_full}. 
For sufficient low $T_J$, 
there are three phases:
$T_S<T_S^{c1}(T_J)$, the phase in which frustration 
remains in spite of adaptation;
$T_S^{c1}(T_J)\leq T_S\leq T_S^{c2}(T_J)$, 
the phase 
giving LMSs; 
$T_S>T_S^{c2}(T_J)$, the phase 
in which no adaptation and frustration is seen.

\begin{figure}
\includegraphics[width=\figwidth]{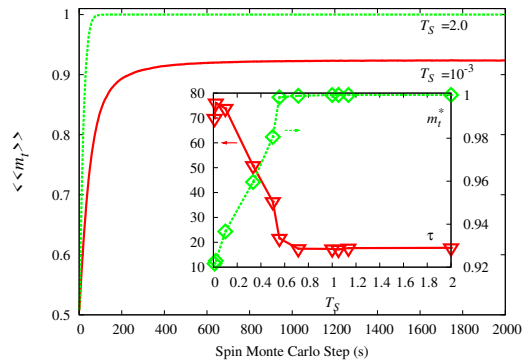}
\caption{(color online) 
Relaxation dynamics of the averaged magnetization of target spins, 
$\langle m_t\rangle_0$ on the adapted interaction
$\bm{J}^{adp}_{T_S}$ for $T_S=10^{-3}$ and $T_S=2.0$. The magnetization $\langle m_t\rangle_0$ is 
evaluated by taking the average over 30 initial conditions for each $\bm{J}^{adp}_{T_S}$ and 1000 different samples of 
$\bm{J}^{adp}_{T_S}$.
The inset shows the dependence of the estimated convergent value of $\langle m_t\rangle_0$ on $T_S$, $m_t^\ast$ 
represents the right axis and the relaxation time $\tau$ represents the left axis. 
} 
\label{Fit_kanwa}
\end{figure}

Let us consider the relaxation dynamics of spins for each $\bm{J}$ 
adapted through evolution under a given $T_S$
and $T_J$, denoted as $\bm{J}^{adp}_{T_S}$, where $T_J$ is fixed at $0.5\times 10^{-3}$.
To understand how the relaxation dynamics depends on $\bm{J}^{adp}_{T_S}$,
instead of using EMC, we adopt standard MC by with the temperature
$T_S^\prime$, fixed at $10^{-5}$
, independently of $T_S$ used in obtaining
$\bm{J}^{adp}_{T_S}$. We compute the temporal change of the target magnetization $m_t=|\sum_{i\in\bm{t}}S_i|$.
Relaxation dynamics of $\langle m_t\rangle_0$ 
for $\bm{J}^{adp}_{T_S}$, $T_S=10^{-3}(\leq T_S^{c1})$, and $T_S=2.0(T_S^{c1}\leq T_S\leq T_S^{c2})$ 
are plotted in \Fref{Fit_kanwa}, 
where $\langle\cdots\rangle_0$ denotes the average over
the randomly chosen initial conditions.
As shown in \Fref{Fit_kanwa}, the relaxation process for $\bm{J}^{adp}_{T_S}$ evolved at low temperatures is much slower. 
Furthermore, $\langle m_t\rangle_0$ 
converges to a value $m_t^\ast$ lower than 1 and remains at that value for a long time.
Depending on the initial condition, the spins are often trapped at a local minimum, 
so that the target configuration is not realized over a long time
span. 

Such dependence on initial conditions is not observed for
$\bm{J}^{adp}_{T_S}$ for $T_S>T_S^{c1}$, where $\langle m_t\rangle_0$ approaches 1 somewhat quickly. 
From an estimate of the convergent value of the target magnetization 
$m_t^\ast$ within the above MC time scale, we obtain the relaxation time $\tau$ 
by fitting to the function $\langle m_t\rangle_0(s)=m_t^{\ast}+c\exp(-s\slash\tau)$,
where $s$ is the MC step of the spin dynamics.
The parameters $m_t^\ast$ and $\tau$ are plotted against $T_S$ in the inset of \Fref{Fit_kanwa}, 
which shows the increase of $\tau$ and the decrease of $m_t^\ast$ from 1 with the decrease of $T_S$ below $T_S^{c1}$. 
These results imply that the energy landscape for the interaction $\bm{J}^{adp}_{T_S}$ is 
rugged for $T_S\leq T_S^{c1}$,
as in a spin-glass phase, 
whereas it is smooth around the target for $T_S^{c1}\leq T_S\leq T_S^{c2}$.
Thus, this landscape is
interpreted as a typical funnel landscape.
It demonstrates a transition from the spin-glass phase to the funnel at $T_S^{c1}$
(see also \cite{SY}).

\begin{figure}
\includegraphics[width=\figwidth]{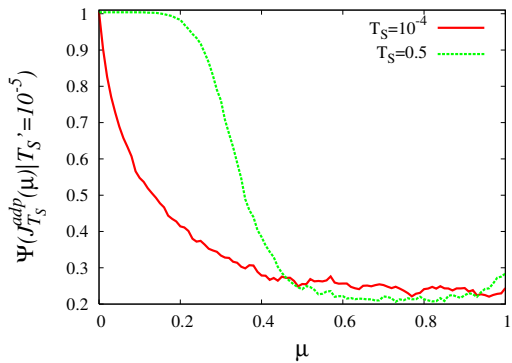}
\caption{(color online) Fitness of the mutated $\bm{J}$ as a function of the mutation rate $\mu$
for $T_S=10^{-4}$ (solid curve) and $T_S=2.0$ (dotted curve). 
For each adapted genotype, the mutated genotypes $\bm{J}$ are generated 150 times
by flipping randomly chosen elements $J_{ij}$ with $\bm{J}$ governed by the rate $\mu$,
and the average is taken over 150 adapted genotypes.}
\label{Fit_vs_mu}
\end{figure}

Now, in the evolved genotypes, let us examine the robustness that 
represents the stability of $\bm{J}$'s fitness with respect to changes 
in the $\bm{J}$ configuration.
From the adopted genotype $\bm{J}^{adp}_{T_S}$,
mutations are imposed by flip-flopping the sign of a certain fraction of 
randomly chosen matrix elements in $\bm{J}^{adp}_{T_S}$. The value of the fraction represents the
mutation rate $\mu$.
We evaluate the fitness of the mutated $\bm{J}^{adp}_{T_S}(\mu)$ at
$T_S^\prime=10^{-5}(\neq T_S)$,
i.e., $\Psi(\bm{J}^{adp}_{T_S}(\mu)|T_S^\prime=10^{-5})$, by taking an
average over 150 samples of mutated $\bm{J}^{adp}_{T_S}(\mu)$.
\Fref{Fit_vs_mu} shows $\mu$ dependence of the fitness 
for $T_S=10^{-4}$ and $T_S=2.0$.
For low values of $T_S$,
the fitness of mutated $\bm{J}^{adp}_{T_S}(\mu)$ exhibits a rapid decrease
with respect to the mutation rate, but
for $T_S$ between $T_S^{c1}$ and $T_S^{c2}$, it does not decrease 
until the mutation rate reaches a specific value.
We define $\mu_c(T_S)$ as the threshold point in the mutation rate beyond which the fitness
$\Psi(\bm{J}^{adp}_{T_S}(\mu)|T_S^\prime)$ begins to decrease from 
unity. \Fref{mu_c} shows the dependence of $\mu_c$ on $T_S$, 
which has a plateau at
$T_S^{c1}\leq T_S\leq T_S^{c2}$ where $\Phi_2$ is unity.
This range of temperatures that exhibits mutational robustness agrees with
the range that gives rise to the LMS. In other words,
mutational robustness is realized
for a genotype with no frustration around the target spins.
Evolution in a mutationally robust genotype $\bm{J}$ is possible
only when the phenotype dynamics is subjected to noise
within the range $T_S^{c1}\leq T_S\leq T_S^{c2}$.
This mutational robustness is interpreted as a consequence that the
fitness landscape becomes non-neutral for $T_S \geq T_S^{c1}$ \cite{SHK_full}.

To check the generality of the transition to the LMS as well as the mutational robustness, 
we have examined the model
by increasing the number of target spins $t$, and confirmed
that the LMSs evolve
at an intermediate range of $T_S$ (that depends on $t$), where both lower energy
and higher fitness are realized together
with mutational robustness, while the actual fitness value therein
decreases with an increase in $t$.
Simulations with larger $N$ (up to 30) have also confirmed
the evolution of the LMSs at an intermediate range of $T_S$ \cite{SHK_full}.

In this study, in order to elucidate the evolutionary origin of robustness
and funnel landscape, we have considered the evolution of a Hamiltonian
system to generate a specific configuration for target spins.
The findings can be summarized as follows.
First, as a result of the formation of a funnel landscape through the
evolution of the Hamiltonian,
robustness to guard against noise is achieved in the dynamic process.
Such shaping of dynamics is possible only
under a certain level of thermal noise, given by temperatures $T_S^{c1}\leq T_S\leq T_S^{c2}$.
Second, under such a temperature range in the process, a funnel-like landscape 
that gives rise to a smooth relaxation dynamics toward the
target phenotype is evolved to avoid the spin-glass phase.
This may explain the ubiquity of such funnel-type 
dynamics observed in evolved biological systems
such as protein folding and gene
expression \cite{KK-PLoS,FangLi-Ouyang-Tang}.
Third, this robustness to thermal noise induces
robustness to mutation; this observation has also been 
discussed for gene transcription network models
\cite{KK-PLoS}. Relevance of thermal noise to robust evolutions is thus demonstrated.

The funnel-like landscape evolved at $T_S^{c1}\leq T_S\leq T_S^{c2}$ 
is characterized by the local
Mattis state without frustration
around the target spins. 
This allows for a smooth and quick relaxation to the target configuration, 
which is in contrast with the relaxation on
a rugged landscape in spin-glass evolved at 
$T_S<T_S^{c1}$, 
where relaxation is often trapped into metastable states.
Theoretical analysis of random spin systems such as replica symmetry
breaking will be relevant to the local Mattis state and mutational
robustness \cite{SHK_full}, as the model discussed in this letter is a variant of spin systems with two temperatures \cite{Sherrington}.

\begin{acknowledgments}
This work was partially supported by a Grant-in-Aid for Scientific Research (No.18079004) from MEXT
and JSPS Fellows (No.20$-$10778) from JSPS. 
\end{acknowledgments}


\begin{thebibliography}{18}
\expandafter\ifx\csname natexlab\endcsname\relax\def\natexlab#1{#1}\fi
\expandafter\ifx\csname bibnamefont\endcsname\relax
  \def\bibnamefont#1{#1}\fi
\expandafter\ifx\csname bibfnamefont\endcsname\relax
  \def\bibfnamefont#1{#1}\fi
\expandafter\ifx\csname citenamefont\endcsname\relax
  \def\citenamefont#1{#1}\fi
\expandafter\ifx\csname url\endcsname\relax
  \def\url#1{\texttt{#1}}\fi
\expandafter\ifx\csname urlprefix\endcsname\relax\def\urlprefix{URL }\fi
\providecommand{\bibinfo}[2]{#2}
\providecommand{\eprint}[2][]{\url{#2}}

\bibitem[{\citenamefont{Waddington}(1957)}]{Waddington}
\bibinfo{author}{\bibfnamefont{C.~H.} \bibnamefont{Waddington}},
  \emph{\bibinfo{title}{The Strategy of the Genes}} (\bibinfo{publisher}{George
  Allen \& Unwin LTD}, \bibinfo{address}{Bristol}, \bibinfo{year}{1957}).

\bibitem[{\citenamefont{Wagner-book}(2005)}]{Wagner-book}
\bibinfo{author}{\bibfnamefont{A.}~\bibnamefont{Wagner}},
  \emph{\bibinfo{title}{Robustness and Evolvability in Living Systems}}
  (\bibinfo{publisher}{Princeton Univ. Pr.},
  \bibinfo{year}{2007}).


\bibitem[{\citenamefont{Alon et~al.}(1999)\citenamefont{Alon, Surette, Barkai,
  and Leibler}}]{Alon}
\bibinfo{author}{\bibfnamefont{U.}~\bibnamefont{Alon} et al. },
  \bibinfo{journal}{Nature} \textbf{\bibinfo{volume}{397}},
  \bibinfo{pages}{168} (\bibinfo{year}{1999}).

\bibitem[{\citenamefont{Go}(1983)}]{Go}
\bibinfo{author}{\bibfnamefont{N.}~\bibnamefont{Go}}, 
\bibinfo{journal}{Ann. Rev. Biophys. Bioeng.}
\textbf{\bibinfo{volume}{12}},
  \bibinfo{pages}{183} (\bibinfo{year}{1983}).


\bibitem[{\citenamefont{Onuchic and Wolynes}(2004)}]{Onuchic}
\bibinfo{author}{\bibfnamefont{J.~N.} \bibnamefont{Onuchic}} \bibnamefont{and}
  \bibinfo{author}{\bibfnamefont{P.~G.} \bibnamefont{Wolynes}},
  \bibinfo{journal}{Curr. Opin. Struc. Biol.}
  \textbf{\bibinfo{volume}{14}}, \bibinfo{pages}{70} (\bibinfo{year}{2004}).

\bibitem[{\citenamefont{Li et~al.}(2004)\citenamefont{Li, Long, Lu, Ouyang, and
  Tang}}]{FangLi-Ouyang-Tang}
\bibinfo{author}{\bibfnamefont{F.}~\bibnamefont{Li} et al. },
  \bibinfo{journal}{Proc. Natl. Acad. Sci. USA} \textbf{\bibinfo{volume}{101}},
  \bibinfo{pages}{4781} (\bibinfo{year}{2004}).

\bibitem[{\citenamefont{Saito et~al.}(1997)\citenamefont{Saito, Sasai, and
  Yomo}}]{SY}
\bibinfo{author}{\bibfnamefont{S.}~\bibnamefont{Saito} et al.}
  \bibinfo{journal}{Proc. Natl. Acad. Sci. USA} \textbf{\bibinfo{volume}{94}},
  \bibinfo{pages}{11324} (\bibinfo{year}{1997}).

\bibitem[{\citenamefont{Ancel and Fontana}(2000)}]{Ancel-Fontana}
\bibinfo{author}{\bibfnamefont{L.~W.} \bibnamefont{Ancel}} \bibnamefont{and}
  \bibinfo{author}{\bibfnamefont{W.}~\bibnamefont{Fontana}},
  \bibinfo{journal}{J. Exp. Zool. (Mol. Dev. Evol.)}
  \textbf{\bibinfo{volume}{288}}, \bibinfo{pages}{242} (\bibinfo{year}{2000}).

\bibitem[{\citenamefont{J. Sun and M. W. Deem}(2007)\citenamefont{J. Sun and M. W. Deem}}]{Deem}
\bibinfo{author}{\bibfnamefont{J.}~\bibnamefont{Sun}}
  \bibnamefont{and} \bibinfo{author}{\bibfnamefont{M.~W.}
	 \bibnamefont{Deem}}, 
 \bibinfo{journal}{Phys. Rev. Lett.} \textbf{\bibinfo{volume}{99}},
  \bibinfo{pages}{228107} (\bibinfo{year}{2007}).


\bibitem[{\citenamefont{Ciliberti et~al.}(2007)\citenamefont{Ciliberti, Martin,
  and Wagner}}]{Wagner}
\bibinfo{author}{\bibfnamefont{S.}~\bibnamefont{Ciliberti} et al.}
  \bibinfo{journal}{PLoS Comput. Biol.} \textbf{\bibinfo{volume}{3}},
  \bibinfo{pages}{165} (\bibinfo{year}{2007}).

\bibitem[{\citenamefont{Kaneko}(2006)}]{KK-book}
\bibinfo{author}{\bibfnamefont{K.}~\bibnamefont{Kaneko}},
  \emph{\bibinfo{title}{Life: An Introduction to Complex Systems Biology}}
  (\bibinfo{publisher}{Springer-Verlag}, \bibinfo{address}{Berlin-New York},
  \bibinfo{year}{2006}).

\bibitem[{\citenamefont{Kaneko}(2007)}]{KK-PLoS}
\bibinfo{author}{\bibfnamefont{K.}~\bibnamefont{Kaneko}},
  \bibinfo{journal}{PLoS ONE} \textbf{\bibinfo{volume}{2}},
  \bibinfo{pages}{e434} (\bibinfo{year}{2007}).

\bibitem[{\citenamefont{Kaneko}(2008)}]{KK-chaos}
\bibinfo{author}{\bibfnamefont{K.}~\bibnamefont{Kaneko}},
  \bibinfo{journal}{Chaos} \textbf{\bibinfo{volume}{18}},
  \bibinfo{pages}{026112} (\bibinfo{year}{2008}).

\bibitem[{\citenamefont{Hukushima and Nemoto}(1996)}]{EMC}
\bibinfo{author}{\bibfnamefont{K.}~\bibnamefont{Hukushima}} \bibnamefont{and}
  \bibinfo{author}{\bibfnamefont{K.}~\bibnamefont{Nemoto}},
  \bibinfo{journal}{J. Phys. Soc. Jpn.} \textbf{\bibinfo{volume}{65}},
  \bibinfo{pages}{1604} (\bibinfo{year}{1996}).

\bibitem[{\citenamefont{Nishimori}(2001)}]{Nishimori}
\bibinfo{author}{\bibfnamefont{H.}~\bibnamefont{Nishimori}},
  \emph{\bibinfo{title}{Statistical Physics of Spin Glasses and Information
  Processing: An Introduction}} (\bibinfo{publisher}{Oxford Univ. Pr.},
  \bibinfo{year}{2001}).


\bibitem{remark} For a finite system with finite $T_J$,
$\Phi_j$ cannot be exactly 1. However, as long as $T_J$ is low,
the deviation from 1 of $\Phi_j$
at the intermediate temperature is negligible.


\bibitem[{\citenamefont{Mattis}(1976)}]{Mattis}
\bibinfo{author}{\bibfnamefont{D.~C.} \bibnamefont{Mattis}},
  \bibinfo{journal}{Phys. Rev.} \textbf{\bibinfo{volume}{56}},
  \bibinfo{pages}{421} (\bibinfo{year}{1976}).



\bibitem[{\citenamefont{Sakata et~al.}(2008)\citenamefont{Sakata, Hukushima,
  and Kaneko}}]{SHK_full}
\bibinfo{author}{\bibfnamefont{A.}~\bibnamefont{Sakata}},
  \bibinfo{author}{\bibfnamefont{K.}~\bibnamefont{Hukushima}},
  \bibnamefont{and} \bibinfo{author}{\bibfnamefont{K.}~\bibnamefont{Kaneko}},
  \bibinfo{note}{unpublished}. 

\bibitem[{\citenamefont{Penney et~al.}(1993)\citenamefont{Penney, Coolen, and
  Sherrington}}]{Sherrington}
\bibinfo{author}{\bibfnamefont{R.~W.}~\bibnamefont{Penney} et al.}
  \bibinfo{journal}{J. Phys. A: Math. Gen.} \textbf{\bibinfo{volume}{26}},
  \bibinfo{pages}{3681} (\bibinfo{year}{1993}).

\end{thebibliography}
\end{document}